

%
%

%
%
%
%

\def\Serif{cmr}
\def\SerifBold{cmbx}
\def\SerifItalics{cmti}
\def\SerifSlanted{cmsl}
\def\SerifBoldItalics{cmbxti}
\def\SansSerif{cmss}
\def\SansSerifBold{cmssbx}
\def\SansSerifItalics{cmssi}
\def\SansSerifSlanted{cmssi}
\def\Math{cmmi}
\def\Symbols{cmsy}
\def\MathBold{cmmib}
\def\MoreSymbols{cmex}
\def\Typewriter{cmtt}
\def\Gothic{eufm}
\def\Double{msbm}
\def\Relazioni{msam}

= 			\Serif10 			at 5pt
= 		\SerifBold10 		at 5pt
= 	\SerifItalics10 	at 5pt
=		\SerifSlanted10 	at 5pt
=	\SerifBoldItalics10	at 5pt
= 		\SansSerif10 		at 5pt
=	\SansSerifBold10	at 5pt
=	\SansSerifItalics10	at 5pt
=	\SansSerifSlanted10	at 5pt
=				\Math10				at 5pt
=			\MathBold10			at 5pt
=			\Symbols10			at 5pt
=		\MoreSymbols10		at 5pt
=		\Typewriter10		at 5pt
=			\Gothic10			at 5pt
=			\Double10			at 5pt

= 			\Serif10 			at 7pt
= 		\SerifBold10 		at 7pt
= 	\SerifItalics10 	at 7pt
=	\SerifSlanted10 	at 7pt
=\SerifBoldItalics10	at 7pt
= 		\SansSerif10 		at 7pt
= 	\SansSerifBold10 	at 7pt
=\SansSerifItalics10	at 7pt
=\SansSerifSlanted10	at 7pt
=			\Math10				at 7pt
=		\MathBold10			at 7pt
=			\Symbols10			at 7pt
=		\MoreSymbols10		at 7pt
=		\Typewriter10		at 7pt
=			\Gothic10			at 7pt
=			\Double10			at 7pt

= 			\Serif10 			at 8pt
= 		\SerifBold10 		at 8pt
= 	\SerifItalics10 	at 8pt
=	\SerifSlanted10 	at 8pt
=\SerifBoldItalics10	at 8pt
= 		\SansSerif10 		at 8pt
= 	\SansSerifBold10 	at 8pt
=\SansSerifItalics10 at 8pt
=\SansSerifSlanted10 at 8pt
=			\Math10				at 8pt
=		\MathBold10			at 8pt
=			\Symbols10			at 8pt
=		\MoreSymbols10		at 8pt
=		\Typewriter10		at 8pt
=			\Gothic10			at 8pt
=			\Double10			at 8pt

= 			\Serif10 			at 10pt
= 		\SerifBold10 		at 10pt
= 		\SerifItalics10 	at 10pt
=		\SerifSlanted10 	at 10pt
=	\SerifBoldItalics10	at 10pt
= 		\SansSerif10 		at 10pt
= 	\SansSerifBold10 	at 10pt
= 	\SansSerifItalics10 at 10pt
= 	\SansSerifSlanted10 at 10pt
=				\Math10				at 10pt
=			\MathBold10			at 10pt
=			\Symbols10			at 10pt
=		\MoreSymbols10		at 10pt
=		\Typewriter10		at 10pt
=			\Gothic10			at 10pt
=			\Double10			at 10pt
=			\Relazioni10			at 10pt

= 				\Serif10 			at 12pt
= 			\SerifBold10 		at 12pt
= 		\SerifItalics10 	at 12pt
=		\SerifSlanted10 	at 12pt
=	\SerifBoldItalics10	at 12pt
= 			\SansSerif10 		at 12pt
= 		\SansSerifBold10 	at 12pt
= 	\SansSerifItalics10 at 12pt
= 	\SansSerifSlanted10 at 12pt
=				\Math10				at 12pt
=			\MathBold10			at 12pt
=			\Symbols10			at 12pt
=		\MoreSymbols10		at 12pt
=			\Typewriter10		at 12pt
=				\Gothic10			at 12pt
=				\Double10			at 12pt

= 			\Serif10 			at 14pt
= 		\SerifBold10 		at 14pt
= 	\SerifItalics10 	at 14pt
=		\SerifSlanted10 	at 14pt
=	\SerifBoldItalics10	at 14pt
= 		\SansSerif10 		at 14pt
= 	\SansSerifBold10 	at 14pt
= \SansSerifSlanted10 at 14pt
= \SansSerifItalics10 at 14pt
=				\Math10				at 14pt
=			\MathBold10			at 14pt
=			\Symbols10			at 14pt
=		\MoreSymbols10		at 14pt
=		\Typewriter10		at 14pt
=			\Gothic10			at 14pt
=			\Double10			at 14pt

\def\NormalStyle{\parindent=5pt\parskip=3pt\normalbaselineskip=14pt%
\def\nt{\tenSerif}%
\def\rm{\fam0\tenSerif}%
\textfont0=\tenSerif\scriptfont0=\sevenSerif\scriptscriptfont0=\fiveSerif
\textfont1=\tenMath\scriptfont1=\sevenMath\scriptscriptfont1=\fiveMath
\textfont2=\tenSymbols\scriptfont2=\sevenSymbols\scriptscriptfont2=\fiveSymbols
\textfont3=\tenMoreSymbols\scriptfont3=\sevenMoreSymbols\scriptscriptfont3=\fiveMoreSymbols
\textfont\itfam=\tenSerifItalics\def\it{\fam\itfam\tenSerifItalics}%
\textfont\slfam=\tenSerifSlanted\def\sl{\fam\slfam\tenSerifSlanted}%
\textfont\ttfam=\tenTypewriter\def\tt{\fam\ttfam\tenTypewriter}%
\textfont\bffam=\tenSerifBold%
\def\bf{\fam\bffam\tenSerifBold}\scriptfont\bffam=\sevenSerifBold\scriptscriptfont\bffam=\fiveSerifBold%
\def\cal{\tenSymbols}%
\def\greekbold{\tenMathBold}%
\def\gothic{\tenGothic}%
\def\Bbb{\tenDouble}%
\def\LieFont{\tenSerifItalics}%
\nt\normalbaselines\baselineskip=14pt%
}

\def\TitleStyle{\parindent=0pt\parskip=0pt\normalbaselineskip=15pt%
\def\nt{\fourteenSansSerifBold}%
\def\rm{\fam0\fourteenSansSerifBold}%
\textfont0=\fourteenSansSerifBold\scriptfont0=\tenSansSerifBold\scriptscriptfont0=\eightSansSerifBold
\textfont1=\fourteenMath\scriptfont1=\tenMath\scriptscriptfont1=\eightMath
\textfont2=\fourteenSymbols\scriptfont2=\tenSymbols\scriptscriptfont2=\eightSymbols
\textfont3=\fourteenMoreSymbols\scriptfont3=\tenMoreSymbols\scriptscriptfont3=\eightMoreSymbols
\textfont\itfam=\fourteenSansSerifItalics\def\it{\fam\itfam\fourteenSansSerifItalics}%
\textfont\slfam=\fourteenSansSerifSlanted\def\sl{\fam\slfam\fourteenSerifSansSlanted}%
\textfont\ttfam=\fourteenTypewriter\def\tt{\fam\ttfam\fourteenTypewriter}%
\textfont\bffam=\fourteenSansSerif%
\def\bf{\fam\bffam\fourteenSansSerif}\scriptfont\bffam=\tenSansSerif\scriptscriptfont\bffam=\eightSansSerif%
\def\cal{\fourteenSymbols}%
\def\greekbold{\fourteenMathBold}%
\def\gothic{\fourteenGothic}%
\def\Bbb{\fourteenDouble}%
\def\LieFont{\fourteenSerifItalics}%
\nt\normalbaselines\baselineskip=15pt%
}

\def\PartStyle{\parindent=0pt\parskip=0pt\normalbaselineskip=15pt%
\def\nt{\fourteenSansSerifBold}%
\def\rm{\fam0\fourteenSansSerifBold}%
\textfont0=\fourteenSansSerifBold\scriptfont0=\tenSansSerifBold\scriptscriptfont0=\eightSansSerifBold
\textfont1=\fourteenMath\scriptfont1=\tenMath\scriptscriptfont1=\eightMath
\textfont2=\fourteenSymbols\scriptfont2=\tenSymbols\scriptscriptfont2=\eightSymbols
\textfont3=\fourteenMoreSymbols\scriptfont3=\tenMoreSymbols\scriptscriptfont3=\eightMoreSymbols
\textfont\itfam=\fourteenSansSerifItalics\def\it{\fam\itfam\fourteenSansSerifItalics}%
\textfont\slfam=\fourteenSansSerifSlanted\def\sl{\fam\slfam\fourteenSerifSansSlanted}%
\textfont\ttfam=\fourteenTypewriter\def\tt{\fam\ttfam\fourteenTypewriter}%
\textfont\bffam=\fourteenSansSerif%
\def\bf{\fam\bffam\fourteenSansSerif}\scriptfont\bffam=\tenSansSerif\scriptscriptfont\bffam=\eightSansSerif%
\def\cal{\fourteenSymbols}%
\def\greekbold{\fourteenMathBold}%
\def\gothic{\fourteenGothic}%
\def\Bbb{\fourteenDouble}%
\def\LieFont{\fourteenSerifItalics}%
\nt\normalbaselines\baselineskip=15pt%
}

\def\ChapterStyle{\parindent=0pt\parskip=0pt\normalbaselineskip=15pt%
\def\nt{\fourteenSansSerifBold}%
\def\rm{\fam0\fourteenSansSerifBold}%
\textfont0=\fourteenSansSerifBold\scriptfont0=\tenSansSerifBold\scriptscriptfont0=\eightSansSerifBold
\textfont1=\fourteenMath\scriptfont1=\tenMath\scriptscriptfont1=\eightMath
\textfont2=\fourteenSymbols\scriptfont2=\tenSymbols\scriptscriptfont2=\eightSymbols
\textfont3=\fourteenMoreSymbols\scriptfont3=\tenMoreSymbols\scriptscriptfont3=\eightMoreSymbols
\textfont\itfam=\fourteenSansSerifItalics\def\it{\fam\itfam\fourteenSansSerifItalics}%
\textfont\slfam=\fourteenSansSerifSlanted\def\sl{\fam\slfam\fourteenSerifSansSlanted}%
\textfont\ttfam=\fourteenTypewriter\def\tt{\fam\ttfam\fourteenTypewriter}%
\textfont\bffam=\fourteenSansSerif%
\def\bf{\fam\bffam\fourteenSansSerif}\scriptfont\bffam=\tenSansSerif\scriptscriptfont\bffam=\eightSansSerif%
\def\cal{\fourteenSymbols}%
\def\greekbold{\fourteenMathBold}%
\def\gothic{\fourteenGothic}%
\def\Bbb{\fourteenDouble}%
\def\LieFont{\fourteenSerifItalics}%
\nt\normalbaselines\baselineskip=15pt%
}

\def\SectionStyle{\parindent=0pt\parskip=0pt\normalbaselineskip=13pt%
\def\nt{\twelveSansSerifBold}%
\def\rm{\fam0\twelveSansSerifBold}%
\textfont0=\twelveSansSerifBold\scriptfont0=\eightSansSerifBold\scriptscriptfont0=\eightSansSerifBold
\textfont1=\twelveMath\scriptfont1=\eightMath\scriptscriptfont1=\eightMath
\textfont2=\twelveSymbols\scriptfont2=\eightSymbols\scriptscriptfont2=\eightSymbols
\textfont3=\twelveMoreSymbols\scriptfont3=\eightMoreSymbols\scriptscriptfont3=\eightMoreSymbols
\textfont\itfam=\twelveSansSerifItalics\def\it{\fam\itfam\twelveSansSerifItalics}%
\textfont\slfam=\twelveSansSerifSlanted\def\sl{\fam\slfam\twelveSerifSansSlanted}%
\textfont\ttfam=\twelveTypewriter\def\tt{\fam\ttfam\twelveTypewriter}%
\textfont\bffam=\twelveSansSerif%
\def\bf{\fam\bffam\twelveSansSerif}\scriptfont\bffam=\eightSansSerif\scriptscriptfont\bffam=\eightSansSerif%
\def\cal{\twelveSymbols}%
\def\bg{\twelveMathBold}%
\def\gothic{\twelveGothic}%
\def\Bbb{\twelveDouble}%
\def\LieFont{\twelveSerifItalics}%
\nt\normalbaselines\baselineskip=13pt%
}

\def\SubSectionStyle{\parindent=0pt\parskip=0pt\normalbaselineskip=13pt%
\def\nt{\twelveSansSerifItalics}%
\def\rm{\fam0\twelveSansSerifItalics}%
\textfont0=\twelveSansSerifItalics\scriptfont0=\eightSansSerifItalics\scriptscriptfont0=\eightSansSerifItalics%
\textfont1=\twelveMath\scriptfont1=\eightMath\scriptscriptfont1=\eightMath%
\textfont2=\twelveSymbols\scriptfont2=\eightSymbols\scriptscriptfont2=\eightSymbols%
\textfont3=\twelveMoreSymbols\scriptfont3=\eightMoreSymbols\scriptscriptfont3=\eightMoreSymbols%
\textfont\itfam=\twelveSansSerif\def\it{\fam\itfam\twelveSansSerif}%
\textfont\slfam=\twelveSansSerifSlanted\def\sl{\fam\slfam\twelveSerifSansSlanted}%
\textfont\ttfam=\twelveTypewriter\def\tt{\fam\ttfam\twelveTypewriter}%
\textfont\bffam=\twelveSansSerifBold%
\def\bf{\fam\bffam\twelveSansSerifBold}\scriptfont\bffam=\eightSansSerifBold\scriptscriptfont\bffam=\eightSansSerifBold%
\def\cal{\twelveSymbols}%
\def\greekbold{\twelveMathBold}%
\def\gothic{\twelveGothic}%
\def\Bbb{\twelveDouble}%
\def\LieFont{\twelveSerifItalics}%
\nt\normalbaselines\baselineskip=13pt%
}

\def\AuthorStyle{\parindent=0pt\parskip=0pt\normalbaselineskip=14pt%
\def\nt{\tenSerif}%
\def\rm{\fam0\tenSerif}%
\textfont0=\tenSerif\scriptfont0=\sevenSerif\scriptscriptfont0=\fiveSerif
\textfont1=\tenMath\scriptfont1=\sevenMath\scriptscriptfont1=\fiveMath
\textfont2=\tenSymbols\scriptfont2=\sevenSymbols\scriptscriptfont2=\fiveSymbols
\textfont3=\tenMoreSymbols\scriptfont3=\sevenMoreSymbols\scriptscriptfont3=\fiveMoreSymbols
\textfont\itfam=\tenSerifItalics\def\it{\fam\itfam\tenSerifItalics}%
\textfont\slfam=\tenSerifSlanted\def\sl{\fam\slfam\tenSerifSlanted}%
\textfont\ttfam=\tenTypewriter\def\tt{\fam\ttfam\tenTypewriter}%
\textfont\bffam=\tenSerifBold%
\def\bf{\fam\bffam\tenSerifBold}\scriptfont\bffam=\sevenSerifBold\scriptscriptfont\bffam=\fiveSerifBold%
\def\cal{\tenSymbols}%
\def\greekbold{\tenMathBold}%
\def\gothic{\tenGothic}%
\def\Bbb{\tenDouble}%
\def\LieFont{\tenSerifItalics}%
\nt\normalbaselines\baselineskip=14pt%
}

\def\AddressStyle{\parindent=0pt\parskip=0pt\normalbaselineskip=14pt%
\def\nt{\eightSerif}%
\def\rm{\fam0\eightSerif}%
\textfont0=\eightSerif\scriptfont0=\sevenSerif\scriptscriptfont0=\fiveSerif
\textfont1=\eightMath\scriptfont1=\sevenMath\scriptscriptfont1=\fiveMath
\textfont2=\eightSymbols\scriptfont2=\sevenSymbols\scriptscriptfont2=\fiveSymbols
\textfont3=\eightMoreSymbols\scriptfont3=\sevenMoreSymbols\scriptscriptfont3=\fiveMoreSymbols
\textfont\itfam=\eightSerifItalics\def\it{\fam\itfam\eightSerifItalics}%
\textfont\slfam=\eightSerifSlanted\def\sl{\fam\slfam\eightSerifSlanted}%
\textfont\ttfam=\eightTypewriter\def\tt{\fam\ttfam\eightTypewriter}%
\textfont\bffam=\eightSerifBold%
\def\bf{\fam\bffam\eightSerifBold}\scriptfont\bffam=\sevenSerifBold\scriptscriptfont\bffam=\fiveSerifBold%
\def\cal{\eightSymbols}%
\def\greekbold{\eightMathBold}%
\def\gothic{\eightGothic}%
\def\Bbb{\eightDouble}%
\def\LieFont{\eightSerifItalics}%
\nt\normalbaselines\baselineskip=14pt%
}

\def\AbstractStyle{\parindent=0pt\parskip=0pt\normalbaselineskip=12pt%
\def\nt{\eightSerif}%
\def\rm{\fam0\eightSerif}%
\textfont0=\eightSerif\scriptfont0=\sevenSerif\scriptscriptfont0=\fiveSerif
\textfont1=\eightMath\scriptfont1=\sevenMath\scriptscriptfont1=\fiveMath
\textfont2=\eightSymbols\scriptfont2=\sevenSymbols\scriptscriptfont2=\fiveSymbols
\textfont3=\eightMoreSymbols\scriptfont3=\sevenMoreSymbols\scriptscriptfont3=\fiveMoreSymbols
\textfont\itfam=\eightSerifItalics\def\it{\fam\itfam\eightSerifItalics}%
\textfont\slfam=\eightSerifSlanted\def\sl{\fam\slfam\eightSerifSlanted}%
\textfont\ttfam=\eightTypewriter\def\tt{\fam\ttfam\eightTypewriter}%
\textfont\bffam=\eightSerifBold%
\def\bf{\fam\bffam\eightSerifBold}\scriptfont\bffam=\sevenSerifBold\scriptscriptfont\bffam=\fiveSerifBold%
\def\cal{\eightSymbols}%
\def\greekbold{\eightMathBold}%
\def\gothic{\eightGothic}%
\def\Bbb{\eightDouble}%
\def\LieFont{\eightSerifItalics}%
\nt\normalbaselines\baselineskip=12pt%
}

\def\RefsStyle{\parindent=0pt\parskip=0pt%
\def\nt{\eightSerif}%
\def\rm{\fam0\eightSerif}%
\textfont0=\eightSerif\scriptfont0=\sevenSerif\scriptscriptfont0=\fiveSerif
\textfont1=\eightMath\scriptfont1=\sevenMath\scriptscriptfont1=\fiveMath
\textfont2=\eightSymbols\scriptfont2=\sevenSymbols\scriptscriptfont2=\fiveSymbols
\textfont3=\eightMoreSymbols\scriptfont3=\sevenMoreSymbols\scriptscriptfont3=\fiveMoreSymbols
\textfont\itfam=\eightSerifItalics\def\it{\fam\itfam\eightSerifItalics}%
\textfont\slfam=\eightSerifSlanted\def\sl{\fam\slfam\eightSerifSlanted}%
\textfont\ttfam=\eightTypewriter\def\tt{\fam\ttfam\eightTypewriter}%
\textfont\bffam=\eightSerifBold%
\def\bf{\fam\bffam\eightSerifBold}\scriptfont\bffam=\sevenSerifBold\scriptscriptfont\bffam=\fiveSerifBold%
\def\cal{\eightSymbols}%
\def\greekbold{\eightMathBold}%
\def\gothic{\eightGothic}%
\def\Bbb{\eightDouble}%
\def\LieFont{\eightSerifItalics}%
\nt\normalbaselines\baselineskip=10pt%
}



%
%


\def\ModeYes{yes}
\def\ModeNo{no}

\def\ModeUndef{undefined}


\def\nx{\noexpand}
\def\ni{\noindent}
\def\newpage{\vfill\eject}

\def\ss{\vskip 5pt}
\def\ms{\vskip 10pt}
\def\bs{\vskip 20pt}

 \def\,{\mskip\thinmuskip}
 \def\!{\mskip-\thinmuskip}
 \def\>{\mskip\medmuskip}
 \def\;{\mskip\thickmuskip}

%
% ... 
%

\def\refsModePost{post}
\def\refsModeAuto{auto}

\def\dbRefsSatusModeOk{ok}
\def\dbRefsSatusModeError{error}
\def\dbRefsSatusModeWarning{warning}


\newcount\BNUM
\BNUM=0

\def\refs{}

\def\SetModePost{\xdef\refsMode{\refsModePost}}			
\SetModePost

\def\dbRefsStatusOk{%
	\xdef\dbRefsStatus{\dbRefsSatusModeOk}%
	\xdef\dbRefsError{\ModeNo}%
	\xdef\dbRefsWarning{\ModeNo}%
	\xdef\dbRefsInfo{\ModeNo}%
}

\def\dbRefs{%
}

\def\dbRefsGet#1{%
	\xdef\found{N}\xdef\ikey{#1}\dbRefsStatusOk%
	\xdef\key{\ModeUndef}\xdef\tag{\ModeUndef}\xdef\tail{\ModeUndef}%
	\dbRefs%
}

\def\NextRefsTag{%
	\global\advance\BNUM by 1%
}
\def\ShowTag#1{{\bf [#1]}}

\def\dbRefsInsert#1#2{%
\dbRefsGet{#1}%
\if\found Y %
   \xdef\dbRefsStatus{\dbRefsSatusModeWarning}%
   \xdef\dbRefsWarning{record is already there}%
   \xdef\dbRefsInfo{record not inserted}%
\else%
   \toks2=\expandafter{\dbRefs}%
   \ifx\refsMode\refsModeAuto \NextRefsTag
    \xdef\dbRefs{%
   	\the\toks2 \nx\xdef\nx\dbx{#1}%
	\nx\ifx\nx\ikey %
		\nx\dbx\nx\xdef\nx\found{Y}%
		\nx\xdef\nx\key{#1}%
		\nx\xdef\nx\tag{\the\BNUM}%
		\nx\xdef\nx\tail{#2}%
	\nx\fi}%
	\global\xdef\refs{\refs \ss\ni[\the\BNUM]\ #2\par}
   \fi%
   \ifx\refsMode\refsModePost 
    \xdef\dbRefs{%
   	\the\toks2 \nx\xdef\nx\dbx{#1}%
	\nx\ifx\nx\ikey %
		\nx\dbx\nx\xdef\nx\found{Y}%
		\nx\xdef\nx\key{#1}%
		\nx\xdef\nx\tag{\ModeUndef}%
		\nx\xdef\nx\tail{#2}%
	\nx\fi}%
   \fi%
\fi%
}

\def\dbRefsEdit#1#2#3{\dbRefsGet{#1}%
\if\found N 
   \xdef\dbRefsStatus{\dbRefsSatusModeError}%
   \xdef\dbRefsError{record is not there}%
   \xdef\dbRefsInfo{record not edited}%
\else%
   \toks2=\expandafter{\dbRefs}%
   \xdef\dbRefs{\the\toks2%
   \nx\xdef\nx\dbx{#1}%
   \nx\ifx\nx\ikey\nx\dbx %
	\nx\xdef\nx\found{Y}%
	\nx\xdef\nx\key{#1}%
	\nx\xdef\nx\tag{#2}%
	\nx\xdef\nx\tail{#3}%
   \nx\fi}%
\fi%
}

\def\bib#1#2{\RefsStyle\dbRefsInsert{#1}{#2}%
	\ifx\dbRefsStatus\dbRefsSatusModeWarning %
		\message{^^J}%
		\message{WARNING: Reference [#1] is doubled.^^J}%
	\fi%
}

\def\ref#1{\dbRefsGet{#1}%
\ifx\found N %
  \message{^^J}%
  \message{ERROR: Reference [#1] unknown.^^J}%
  \ShowTag{??}%
\else%
	\ifx\tag\ModeUndef \NextRefsTag%
		\dbRefsEdit{#1}{\the\BNUM}{\tail}%
		\dbRefsGet{#1}%
		\global\xdef\refs{\refs \ss\ni [\tag]\ \tail\par}
	\fi
	\ShowTag{\tag}%
\fi%
}

\def\ShowBiblio{\ms\Ensure{\SectionEnsure}%
{\SectionStyle\ni References}%
{\RefsStyle\refs}%
}

\newcount\CHANGES
\CHANGES=0
\def\AuxFile{7}
\def\PreventDoubleOn{\xdef\PreventDoubleLabel{\ModeYes}}

\PreventDoubleOn

\def\StoreLabel#1#2{\xdef\itag{#2}
 \ifx\PreModeStatus\ModeNo %
   \message{^^J}%
   \errmessage{You can't use Check without starting with OpenPreMode (and finishing with ClosePreMode)^^J}%
 \else%
   \immediate\write\AuxFile{\nx\dbLabelPreInsert{#1}{\itag}}%
   \dbLabelGet{#1}%
   \ifx\itag\tag %
   \else%
	\global\advance\CHANGES by 1%
 	\xdef\itag{(?.??)}%
    \fi%
   \fi%
}

\def\PreModeStatus{\ModeNo}

\def\ClosePreMode{\immediate\closeout\AuxFile%
  \ifnum\CHANGES=0%
	\message{^^J}%
	\message{**********************************^^J}%
	\message{**  NO CHANGES TO THE AuxFile  **^^J}%
	\message{**********************************^^J}%
 \else%
	\message{^^J}%
	\message{**************************************************^^J}%
	\message{**  PLAEASE TYPESET IT AGAIN (\the\CHANGES)  **^^J}%
    \errmessage{**************************************************^^ J}%
  \fi%
  \edef\PreModeStatus{\ModeNo}%
}

\def\dbLabelSatusModeOk{ok}

\def\dbLabelSatusModeWarning{warning}

\def\dbLabelStatusOk{%
	\xdef\dbLabelStatus{\dbLabelSatusModeOk}%
	\xdef\dbLabelError{\ModeNo}%
	\xdef\dbLabelWarning{\ModeNo}%
	\xdef\dbLabelInfo{\ModeNo}%
}

\def\dbLabel{%
}

\def\dbLabelGet#1{%
	\xdef\found{N}\xdef\ikey{#1}\dbLabelStatusOk%
	\xdef\key{\ModeUndef}\xdef\tag{\ModeUndef}\xdef\pre{\ModeUndef}%
	\dbLabel%
}

\def\ShowLabel#1{%
 \dbLabelGet{#1}%
 \ifx\tag \ModeUndef %
 	\global\advance\CHANGES by 1%
 	(?.??)%
 \else%
 	\tag%
 \fi%
}

\def\dbLabelPreInsert#1#2{\dbLabelGet{#1}%
\if\found Y %
  \xdef\dbLabelStatus{\dbLabelSatusModeWarning}%
   \xdef\dbLabelWarning{Label is already there}%
   \xdef\dbLabelInfo{Label not inserted}%
   \message{^^J}%
   \errmessage{Double pre definition of label [#1]^^J}%
\else%
   \toks2=\expandafter{\dbLabel}%
    \xdef\dbLabel{%
   	\the\toks2 \nx\xdef\nx\dbx{#1}%
	\nx\ifx\nx\ikey %
		\nx\dbx\nx\xdef\nx\found{Y}%
		\nx\xdef\nx\key{#1}%
		\nx\xdef\nx\tag{#2}%
		\nx\xdef\nx\pre{\ModeYes}%
	\nx\fi}%
\fi%
}

\def\dbLabelInsert#1#2{\dbLabelGet{#1}%
\xdef\itag{#2}%
\dbLabelGet{#1}%
\if\found Y %
	\ifx\tag\itag %
	\else%
	   \ifx\PreventDoubleLabel\ModeYes %
		\message{^^J}%
		\errmessage{Double definition of label [#1]^^J}%
	   \else%
		\message{^^J}%
		\message{Double definition of label [#1]^^J}%
	   \fi%
	\fi%
   \xdef\dbLabelStatus{\dbLabelSatusModeWarning}%
   \xdef\dbLabelWarning{Label is already there}%
   \xdef\dbLabelInfo{Label not inserted}%
\else%
   \toks2=\expandafter{\dbLabel}%
    \xdef\dbLabel{%
   	\the\toks2 \nx\xdef\nx\dbx{#1}%
	\nx\ifx\nx\ikey %
		\nx\dbx\nx\xdef\nx\found{Y}%
		\nx\xdef\nx\key{#1}%
		\nx\xdef\nx\tag{#2}%
		\nx\xdef\nx\pre{\ModeNo}%
	\nx\fi}%
\fi%
}


\newcount\PART
\newcount\CHAPTER
\newcount\SECTION
\newcount\SUBSECTION
\newcount\FNUMBER

\PART=0
\CHAPTER=0
\SECTION=0
\SUBSECTION=0	
\FNUMBER=0

\def\LastPart{\ModeUndef}
\def\LastChapter{\ModeUndef}
\def\LastSection{\ModeUndef}
\def\LastSubSection{\ModeUndef}
\def\LastClaim{\ModeUndef}
\def\Last{\ModeUndef}

\newdimen\TOBOTTOM
\newdimen\LIMIT

\def\Ensure#1{\ \par\ \immediate\LIMIT=#1\immediate\TOBOTTOM=\the\pagegoal\advance\TOBOTTOM by -\pagetotal%
\ifdim\TOBOTTOM<\LIMIT\newpage \else%
\vskip-\parskip\vskip-\parskip\vskip-\baselineskip\fi}

\def\PartLabel{\the\PART}
\def\NewPart#1{\global\advance\PART by 1%
         \bs\ni{\PartStyle  Part \PartLabel:}
         \bs\ni{\PartStyle #1}\newpage%
         \CHAPTER=0\SECTION=0\SUBSECTION=0\FNUMBER=0%
         \gdef\Left{#1}%
         \global\edef\Last{\PartLabel}%
         \global\edef\LastPart{\PartLabel}%
         \global\edef\LastChapter{\ModeUndef}%
         \global\edef\LastSection{\ModeUndef}%
         \global\edef\LastSubSection{\ModeUndef}%
         \global\edef\LastClaim{\ModeUndef}}
\def\ChapterLabel{\the\CHAPTER}
\def\NewChapter#1{\global\advance\CHAPTER by 1%
         \bs\ni{\ChapterStyle  Chapter \ChapterLabel: #1}\ms%
         \SECTION=0\SUBSECTION=0\FNUMBER=0%
         \gdef\Left{#1}%
         \global\edef\Last{\ChapterLabel}%
         \global\edef\LastChapter{\ChapterLabel}%
         \global\edef\LastSection{\ModeUndef}%
         \global\edef\LastSubSection{\ModeUndef}%
         \global\edef\LastClaim{\ModeUndef}}
\def\SectionEnsure{3cm}
\def\NewSection#1{\Ensure{\SectionEnsure}\gdef\SectionLabel{\the\SECTION}\global\advance\SECTION by 1%
         \ms\ni{\SectionStyle  \SectionLabel.\ #1}\ss%
         \SUBSECTION=0\FNUMBER=0%
         \gdef\Left{#1}%
         \global\edef\Last{\SectionLabel}%
         \global\edef\LastSection{\SectionLabel}%
         \global\edef\LastSubSection{\ModeUndef}%
         \global\edef\LastClaim{\ModeUndef}}
\def\NewAppendix#1#2{\Ensure{\SectionEnsure}\gdef\SectionLabel{#1}\global\advance\SECTION by 1%
         \bs\ni{\SectionStyle  Appendix \SectionLabel.\ #2}\ss%
         \SUBSECTION=0\FNUMBER=0%
         \gdef\Left{#2}%
         \global\edef\Last{\SectionLabel}%
         \global\edef\LastSection{\SectionLabel}%
         \global\edef\LastSubSection{\ModeUndef}%
         \global\edef\LastClaim{\ModeUndef}}
\def\Acknowledgements{\Ensure{\SectionEnsure}\gdef\SectionLabel{}%
         \ms\ni{\SectionStyle  Acknowledgments}\ss%
         \SECTION=0\SUBSECTION=0\FNUMBER=0%
         \gdef\Left{}%
         \global\edef\Last{\ModeUndef}%
         \global\edef\LastSection{\ModeUndef}%
         \global\edef\LastSubSection{\ModeUndef}%
         \global\edef\LastClaim{\ModeUndef}}
\def\SubSectionEnsure{2cm}
\def\SubSectionLabel{\ifnum\SECTION>0 \the\SECTION.\fi\the\SUBSECTION}
\def\NewSubSection#1{\Ensure{\SubSectionEnsure}\global\advance\SUBSECTION by 1%
         \ms\ni{\SubSectionStyle #1}\ss%
         \global\edef\Last{\SubSectionLabel}%
         \global\edef\LastSubSection{\SubSectionLabel}}
\def\SetNumberingModeN{\def\ClaimLabel{(\the\FNUMBER)}}
\def\SetNumberingModeSN{\def\ClaimLabel{(\ifnum\SECTION>0 \SectionLabel.\fi%
      \the\FNUMBER)}}
\def\SetNumberingModeCSN{\def\ClaimLabel{(\ifnum\CHAPTER>0 \the\CHAPTER.\fi%
      \ifnum\SECTION>0 \SectionLabel.\fi%
      \the\FNUMBER)}}

\def\NewClaim{\global\advance\FNUMBER by 1%
    \ClaimLabel%
    \global\edef\LastClaim{\ClaimLabel}%
    \global\edef\Last{\ClaimLabel}}

\def\HideLabels{\xdef\ShowLabelsMode{\ModeNo}}
\HideLabels

\def\fn{\eqno{\NewClaim}} 
\def\fl#1{%
\ifx\ShowLabelsMode\ModeYes%
 \eqno{{\buildrel{\hbox{\AbstractStyle[#1]}}\over{\hfill\NewClaim}}}%
\else%
 \eqno{\NewClaim}%
\fi%
\dbLabelInsert{#1}{\ClaimLabel}}
\def\fprel#1{\global\advance\FNUMBER by 1\StoreLabel{#1}{\ClaimLabel}%
\ifx\ShowLabelsMode\ModeYes%
\eqno{{\buildrel{\hbox{\AbstractStyle[#1]}}\over{\hfill.\itag}}}%
\else%
 \eqno{\itag}%
\fi%
}

\def\cl#1{\global\advance\FNUMBER by 1\dbLabelInsert{#1}{\ClaimLabel}%
\ifx\ShowLabelsMode\ModeYes%
${\buildrel{\hbox{\AbstractStyle[#1]}}\over{\hfill\ClaimLabel}}$%
\else%
  $\ClaimLabel$%
\fi%
}
\def\cprel#1{\global\advance\FNUMBER by 1\StoreLabel{#1}{\ClaimLabel}%
\ifx\ShowLabelsMode\ModeYes%
${\buildrel{\hbox{\AbstractStyle[#1]}}\over{\hfill.\itag}}$%
\else%
  $\itag$%
\fi%
}

\def\Note{\ms\leftskip 3cm\rightskip 1.5cm\AbstractStyle}
\def\endNote{\par\leftskip 2cm\rightskip 0cm\NormalStyle\ss}


\parindent=7pt
\leftskip=2cm
\newcount\SideIndent
\newcount\SideIndentTemp
\SideIndent=0
\newdimen\SectionIndent
\SectionIndent=-8pt

\def\sidebar{\vrule height15pt width.2pt }
\def\endcorner{\hbox{\hbox{\vrule height6pt width.2pt}\vbox to6pt{\vfill\hbox
to4pt{\leaders\hrule height0.2pt\hfill}}}}
\def\begincorner{\hbox{\hbox{\vrule height6pt width.2pt}\vbox to6pt{\hbox
to4pt{\leaders\hrule height0.2pt\hfill}}}}
\def\endbegincorner{\hbox{\vbox to15pt{\endcorner\vskip-6pt\begincorner\vfill}}}
\def\SideShow{\SideIndentTemp=\SideIndent \ifnum \SideIndentTemp>0 
\loop\sidebar\hskip 2pt \advance\SideIndentTemp by-1\ifnum \SideIndentTemp>1 \repeat\fi}

\def\BeginSection{{\vbadness 100000 \par\ni\hskip\SectionIndent%
\SideShow\vbox to 15pt{\vfill\begincorner}}\global\advance\SideIndent by1\vskip-10pt}

\def\EndSection{{\vbadness 100000 \par\ni\global\advance\SideIndent by-1%
\hskip\SectionIndent\SideShow\vbox to15pt{\endcorner\vfill}\vskip-10pt}}

\def\EndBeginSection{{\vbadness 100000\par\ni%
\global\advance\SideIndent by-1\hskip\SectionIndent\SideShow
\vbox to15pt{\vfill\endbegincorner}}%
\global\advance\SideIndent by1\vskip-10pt}

\def\ShowBeginCorners#1{%
\SideIndentTemp =#1 \advance\SideIndentTemp by-1%
\ifnum \SideIndentTemp>0 %
\vskip-15truept\hbox{\kern 2truept\vbox{\hbox{\begincorner}%
\ShowBeginCorners{\SideIndentTemp}\vskip-3truept}}%
\fi%
}

\def\ShowEndCorners#1{%
\SideIndentTemp =#1 \advance\SideIndentTemp by-1%
\ifnum \SideIndentTemp>0 %
\vskip-15truept\hbox{\kern 2truept\vbox{\hbox{\endcorner}%
\ShowEndCorners{\SideIndentTemp}\vskip 2truept}}%
\fi%
}

\def\BeginSections#1{{\vbadness 100000 \par\ni\hskip\SectionIndent%
\SideShow\vbox to 15pt{\vfill\ShowBeginCorners{#1}}}\global\advance\SideIndent by#1\vskip-10pt}

\def\EndSections#1{{\vbadness 100000 \par\ni\global\advance\SideIndent by-#1%
\hskip\SectionIndent\SideShow\vbox to15pt{\vskip15pt\ShowEndCorners{#1}\vfill}\vskip-10pt}}

\def\EndBeginSections#1#2{{\vbadness 100000\par\ni%
\global\advance\SideIndent by-#1%
\hbox{\hskip\SectionIndent\SideShow\kern-2pt%
\vbox to15pt{\vskip15pt\ShowEndCorners{#1}\vskip4pt\ShowBeginCorners{#2}}}}%
\global\advance\SideIndent by#2\vskip-10pt}




%
%


\def\al{\alpha}
\def\be{\beta}
\def\de{\delta}

\def\ep{\epsilon}

\def\te{\theta}
\def\la{\lambda}
\def\ze{\zeta}
\def\om{\omega}
\def\si{\sigma}
\def\vp{\varphi}

\def\Ga{\Gamma}

\def\Si{\Sigma}


 \def\calS{{\hbox{\cal S}}}




 \def\R{{\hbox{\Bbb R}}}

 \def\R{{\hbox{\Bbb R}}}


\def\GL{{\hbox{GL}}}
\def\det{{\hbox{det}}}

\def\Lor{{\hbox{Lor}}}
\def\Diff{{\hbox{Diff}}}

\def\ip{\hbox to4pt{\leaders\hrule height0.3pt\hfill}\vbox to8pt{\leaders\vrule width0.3pt\vfill}\kern 2pt}
 
\def\del{\partial}
\def\na{\nabla}

\def\Vec{\hbox{\gothic X}}

\def\arr{\rightarrow}

%
%

\NormalStyle
\SetNumberingModeSN
\PreventDoubleOn

\long\def\title#1{\centerline{\TitleStyle\ni#1}}

\long\def\author#1{\ms\centerline{\AuthorStyle by {\it #1}}}

\long\def\address#1{\ss\centerline{\AddressStyle #1}\par}
\long\def\moreaddress#1{\centerline{\AddressStyle #1}\par}
\def\abstract{\ms\leftskip 3cm\rightskip .5cm\AbstractStyle{\bf \ni Abstract:}\ }
\def\endabstract{\par\leftskip 2cm\rightskip 0cm\NormalStyle\ss}

\SetNumberingModeSN

\def\calG{{\hbox{\cal G}}}

\def\calD{{\hbox{\cal D}}}

\def\frac[#1/#2]{\hbox{$#1\over#2$}}
\def\Frac[#1/#2]{{#1\over#2}}
\def\({\left(}
\def\){\right)}
\def\[{\left[}
\def\]{\right]}
\def\^#1{{}^{#1}_{\>\cdot}}
\def\_#1{{}_{#1}^{\>\cdot}}
\def\Label=#1{{\buildrel {\hbox{\fiveSerif \ShowLabel{#1}}}\over =}}
\def\<{\kern -1pt}

\def\Dal{\hbox{\tenRelazioni  \char003}}


\def\ExpandAllCNotes{\long\def\CNote##1{%
\BeginSection
	\Note%
 		##1%
	\endNote%
\EndSection%
}}
\ExpandAllCNotes
%
%
%
%


\def\frame#1{\vbox{\hrule\hbox{\vrule\vbox{\kern2pt\hbox{\kern2pt#1\kern2pt}\kern2pt}\vrule}\hrule\kern-4pt}}

\def\Box to #1#2#3{\frame{\vtop{\hbox to #1{\hfill #2 \hfill}\hbox to #1{\hfill #3 \hfill}}}}


\bib{Marta1}{M.Campigotto, L.Fatibene,
{\it Gauge Natural Formulation of Conformal Theory of Gravity},
Annals of Phys.,  354, 328 (2015); 
arXiv:1404.0898 [gr-qc]
}

\bib{Marta2}{M.Campigotto, L.Fatibene,
{\it Generally Covariant vs. Gauge Structure for Conformal Field Theories},
Annals Phys., 362 (2015) 521-528;
arXiv:1506.06071 [gr-qc]
}

\bib{Jackiw}{R. Jackiw, So-Young Pi, 
{\it Fake Conformal Symmetry in Conformal Cosmological Models}, 
Phys. Rev. D 91, 067501; arXiv:1407.8545 [gr-qc]}

\bib{tHooft}{G. 't Hooft, 
{\it Local Conformal Symmetry: the Missing Symmetry Component for Space and Time}, 
arXiv:1410.6675 [gr-qc], 2014}

\bib{Manheim}{ P.D. Mannheim, D.Kazanas, 
{\it Exact Vacuum Solution to Conformal Weyl Gravity and Galactic Rotation Curves}, 
Astrophys. J. 342 (1989), pp. 635?638.}

\bib{Augmented}{L. Fatibene, M. Ferraris, M. Francaviglia, 
{\it Augmented Variational Principles and Relative Conservation Laws in Classical Field Theory}, 
International Journal of Geometric Methods in Modern Physics 2 (June 2005), pp.
373?392; arXiv:math-ph/0411029.}

\bib{Taub}{ R. Clarkson, L. Fatibene, R.B. Mann, Nuclear Phys. B 652(1-3) (2003), 348?382.}

\bib{Wald}{ L. Fatibene, M. Ferraris, M. Francaviglia, M. Raiteri, 
{\it Remarks on N?other charges and black holes entropy}, 
Ann. Physics 275 (1999), no. 1, 2753}

\bib{Libro}{L. Fatibene and M. Francaviglia, 
{\it Natural and Gauge Natural Formalism for Classical Field Theories}, 
Kluver Academic Publisher, Dordrecht, 2003}

\bib{Katz}{ J. Katz, Class. Quantum Grav., 2, 1985, 423}

\bib{PhysState}{ L.Fatibene, M.Ferraris, G.Magnano, 
{\it Constraining the Physical State by Symmetries}, 
(in preparation)}

\bib{Eck}{D.J. Eck, 
{\it Gauge-natural bundles and generalized gauge theories}, 
Mem. Amer. Math. Soc. 33(247) (1981)}

\bib{Kolar}{I. Kolar, P. W. Michor, J. Slovak, 
{\it Natural Operations in Differential Geometry,}
Springer- Verlag, New York (1993)}

\bib{Norton}{J.D. Norton, 
{\it General covariance, gauge theories and the Kretschmann objection},
in: Katherine Brading and Elena Castellani (eds.), 
{\it Symmetries in Physics: Philosophical Reflections}. 
Cambridge University Press. 110?123 (2003)} 

\bib{Stachel}{M. Iftime, J.Stackel, 
{\it The Hole Argument for Covariant Theories}, 
Gen.Rel.Grav. 38 (2006) 1241?1252}

\bib{ADM}{R. Arnowitt, S. Deser and C. W. Misner, in: 
{\it Gravitation: An Introduction to Current Research}, 
L. Witten ed. Wyley, 227, (New York, 1962); gr-qc/0405109}



\def\ubal{\underline{\al}\kern1pt}
\def\obal{\overline{\al}\kern1pt}

\def\ubR{\underline{R}\kern1pt}
\def\obR{\overline{R}\kern1pt}
\def\ubom{\underline{\om}\kern1pt}
\def\obxi{\overline{\xi}\kern1pt}
\def\ubu{\underline{u}\kern1pt}
\def\ube{\underline{e}\kern1pt}
\def\obe{\overline{e}\kern1pt}

\NormalStyle

\title{Conformal Gravity as a Gauge Natural Theory}

\author{ M.Campigotto$^{a,b}$, L.Fatibene$^{b, c}$}

\address{$^a$ Department of Physics, University of Torino (Italy)}
\moreaddress{$^b$ INFN - Sezione Torino}
\moreaddress{$^c$ Department of Mathematics, University of Torino (Italy)}

\abstract
We shall review conformal gravity as a gauge natural theory and discuss the consequences of Weyl covariance on the definition of physical states.
\endabstract

\NewSection{Introduction}

In the literature there are two types of transformations which are often called {\it conformal transformations}.

The first class of transformations are maps in a Riemannian manifold $(M, g)$ which preserve angles or, equivalently, 
they preserve inner products up to a scalar field $\vp$, which is called the {\it conformal factor}, namely
 diffeomorphisms $\Phi\in \Diff(M)$ such that
$$
(\Phi^\ast g)(v, w)= g(\Phi_\ast v, \Phi_\ast w)= \vp(x)\cdot g(v, w)
\fn$$
for any two vector fields $v, w\in \Vec(M)$.

The second class of transformations, which we call {\it Weyl tranformations} are gauge transformations which change the metric to a different metric $\tilde g= \vp \cdot g$, without affecting the spacetime point.
These are vertical transformations on the configuration bundle $\Lor(M)$ of Lorentzian metrics on $M$.

The first class is often used  in conformal theories which are field theories on Minkowski spacetime which are covariant with respect to a bigger group than the Poincar\'e group which includes conformal transformations.
The second class is used  to consider generally covariant theories with an extra gauge symmetry corresponding to the Weyl transformations.
Of course, there is no point in considering conformal transformations in a generally covariant theory which is already covariant with respect to any spacetime diffeomorphism, hence including conformal transformations.

In \ref{Marta1} and \ref{Marta2} we discussed superpotentials for the so--called {\it conformal gravity}; see also \ref{Manheim}.
There we found the expression for superpotential of the theory described by the Lagrangian
$$
L_W= a \sqrt{g} \>W^{\al\be\mu\nu}W_{\al\be\mu\nu}= a \sqrt{g}\[ 3R^{\al\be\mu\nu}R_{\al\be\mu\nu} -6 R^{\mu\nu}R_{\mu\nu} + R^2\]
\fl{Lag}$$
where the Weyl tensor $W$ is defined as
$$
W_{\al\be\mu\nu}= R_{\al\be\mu\nu} -\(g_{\al(\mu} R_{\nu]\be} - g_{\be(\mu} R_{\nu]\al}\) +\frac[1/3]R g_{\al[\mu}g_{\nu]\be}
\fn$$

Usually in the literature the Lagrangian \ShowLabel{Lag} is considered in the ``equivalent'' form 
$$
L_W^\ast=  2a \sqrt{g}\[  3 R^{\mu\nu}R_{\mu\nu} - R^2\]= L_W-3a G
\fl{Lag2}$$
which is obtained by subtracting a Gauss-Bonnet term 
$$
G= \sqrt{g}\[ R^{\al\be\mu\nu}R_{\al\be\mu\nu} -4 R^{\mu\nu}R_{\mu\nu} + R^2\]
\fn$$

Although the Gauss-Bonnet term is known to be a local divergence and as such it does not affect field equations,
neglecting it is a particularly bad idea, since the Gauss-Bonnet term is generally covariant, but not Weyl covariant.
Accordingly, while total Lagrangian $L_W$ is Weyl covariant, the ``simplified'' Lagrangian $L_W^\ast$ is not, 
which is not a good thing if one wants to study conservation laws.

About Weyl covariant theories there was recently a discussion about the physical content of the gauge covariance with respect to Weyl transformations; see \ref{Jackiw}, \ref{tHooft}.
We in fact confirmed what they argued in a simpler example, that Noether currents induced by Weyl transformations for the Lagrangian \ShowLabel{Lag} are identically zero off-shell.
Accordingly, as argued in  \ref{Jackiw} they fail to define gauge transformations in Hamiltonian framework so that Weyl transformations do not seem to play a role for field equations, for conservation laws or for the definition of the physical state.

\ms
Hereafter we shall review Weyl covariant theories within the framework of gauge natural theories; see \ref{Libro}.
This is important since within the framework of gauge natural theories one has a standard treatment of conservation laws
(see \ref{Augmented}) which has proven to be able to solve a number of issues traditionally connected to conservation laws in generally covariant theories (e.g.~the anomalous factor; see \ref{Katz}) as well as to connect to recent important issues (e.g.~the definition of entropy of gravitational system; see \ref{Wald}, \ref{Taub} and references quoted therein).

The purpose of this paper is to discuss within this framework the issue related to the physical state; see also \ref{PhysState}.
In particular we shall argue that even when gauge symmetries do not give information in field equations or conservation laws
they still constrain the definition of physical states of the theory by considering different configurations as physically equivalent.

\NewSection{Gauge natural setting for conformal gravity}

The action of Weyl transformations and spacetime diffeomorphisms is captured by the  action
$$
\la: \GL(m)\times \R\times L\arr L: (J, l, g_{ab}),\mapsto g'_{ab}= \exp(l) \bar J^c_a g_{cd} \bar J^d_b
\fl{GroupAction}$$
of the group $\GL(m)\times \R$ on the set $L$ of non-degenerate forms of Lorentzian signature $\eta=(3, 1)$.

If we wish to see Weyl transformations as gauge transformations one should define a structure bundle $P$ to be a $\R$-principal bundle and use the action \ShowLabel{GroupAction} to define an associated gauge natural bundle, namely
$$
C= \(L(M)\times P\) \times_{ \la} L
\fn$$
where $L(M)$ denotes the general frame bundle; see \ref{Eck}, \ref{Kolar} for gauge natural bundles.

For a pair coordinate systems $(x^\mu, e_a^\mu)$ on the frame bundle $L(M)$ and $(x^\mu, l)$ on $P$ one can define
fibered coordinates $(x^\mu, g_{\mu\nu})$ on $C$ by setting $g_{\mu\nu}:= \exp(l)\> e_\mu^a g_{ab} e^b_\nu$.

If we change coordinates on the structure bundle
$$
x'^\mu=x'^\mu(x)
\qquad
l'= \om(x) + l
\qquad
e'^\mu_a= J^\mu_\nu e^\nu_a
\fl{TransFunctions}$$
the coordinates on $C$ change accordingly as
$$
x'^\mu=x'^\mu(x)
\qquad
g'_{\mu\nu} = e^{\om(x)} \bar J^\al_\mu g_{\al\be} \bar J^\be_\nu
\fn$$
As one sees from equation \ShowLabel{TransFunctions} transition functions on $P$ are affine transformations on the fiber $\R$;
accordingly, the bundle $P$ is principal and affine at the same time.
Being it affine it allows global sections, being it a principal with global sections it is trivial.
This holds true for any $\R$-principal bundle; any $\R$-principal bundle is necessarily trivial.
As a consequence any two $\R$-principal bundles are isomorphic being both trivial.

An automorphism on $P$ is locally described as
$$
x'^\mu=x'^\mu(x)
\qquad
l'= \om(x) + l
\fl{GaugeTransf}$$
This induces an automorphism on $L(M)$ by natural lift
$$
x'^\mu=x'^\mu(x)
\qquad
e'^\mu_a= J^\mu_\nu e^\nu_a
\fn$$
and one on the associated bundle $C$
$$
x'^\mu=x'^\mu(x)
\qquad
g'_{\mu\nu} = e^{\om(x)} \bar J^\al_\mu g_{\al\be} \bar J^\be_\nu
\fn$$
which in fact combines a diffeomorphism and a Weyl transformation.

One--parameter flows of automorphisms on $P$ are described by right invariant vector fields
$$
\Xi = \xi^{\mu}(x) \Frac[\del/\del x^\mu] + \ze(x)\rho
\qquad
\rho=\Frac[\del/\del l]
\fn$$ 
 where $\rho$ is a right invariant pointwise basis for vertical vectors on $P$.
 They induce a one parameter flow on $C$ which is generated by vector fields in the form
 $$
 \Xi_\la= \xi^{\mu}(x) \Frac[\del/\del x^\mu] -  \(\del_\mu \xi^\al g_{\al\nu}+\del_\nu \xi^\al g_{\al\mu} - \ze g_{\mu\nu}\)  \Frac[\del/\del g_{\mu\nu}]
 \fn$$

Given a manifold $M$ one can choose an atlas with transition functions $J^\mu_\nu$ which are the Jacobians of coordinate changes.
Given a Jacobian one can define a group homomorphism
$$
i: \GL(m) \arr \R: J\mapsto \ln(\det(J))
\fn$$
Then one can define a natural bundle $P(M)= L(M)\times _i \R$ which is $\R$-principal too.
Since, as we discussed above, there exists only one $\R$-principal bundle up to bundle isomorphisms,
then any $\R$-principal bundle  $P$ is in fact trivial and isomorphic to $P(M)$ and hence
 any $\R$-principal bundle  $P$ is in fact natural.
 
Given a (torsionless) connection $\Ga^\al_{\be\mu}$ on $M$, then one can define the coefficients $\Ga_\mu=\Ga^\al_{\al\mu}$ 
which transform as
$$
\Ga'_{\mu}=\Ga'^\al_{\al\mu} = J^\al_\ep \( \Ga^\ep_{\rho\si} \bar J^\rho_\al \bar J^\si_\mu + \bar J^\ep_{\al\mu}\)
=  \bar J^\si_\mu\(  \Ga_{\si}   - d_\si \ln(J) \)
\fn$$
They accordingly define a principal connection on $P$, namely
$$
\te= dx^\mu\otimes \(\del_\mu - \Ga_\mu \rho\)
\fn$$
Then any symmetry generator $\Xi$ can be split into a horizontal and a vertical part
$$
\Xi= \xi^{\mu} \(\del_\mu - \Ga_\mu\rho\)\oplus (\ze + \Ga_\mu \xi^\mu)\rho= \Xi_H\oplus \Xi_V
\qquad(\ze_V= \ze + \Ga_\mu \xi^\mu)
\fn$$
as well as into a natural and a vertical part
$$
\Xi= \( \xi^{\mu}\del_\mu + \del_\al\xi^\al\rho\) \oplus \(\ze - \del_\al\xi^\al\)\rho
\fl{XiSplit}$$
Let us stress that the second decomposition, namely \ShowLabel{XiSplit}, has nothing to do with a connection and
it is canonical in view of the canonical natural structure on $P(M)$. 
It splits the one parameter flow of transformations \ShowLabel{GaugeTransf} into a flow of diffeomorphisms and a flow of Weyl transformations, this splitting being canonical and global.

The splitting is also induced on $C$ obtaining
$$
 \Xi_\la= \( \xi^\mu \del_\mu  -  \(\del_\mu \xi^\al g_{\al\nu}+\del_\nu \xi^\al g_{\al\mu} -\del_\al\xi^\al g_{\mu\nu}\)  \Frac[\del/\del g_{\mu\nu}]\)
 \oplus \( \ze - \del_\al\xi^\al\) g_{\mu\nu}  \Frac[\del/\del g_{\mu\nu}] = \hat \xi \oplus \Xi_W
\fn$$

In fact one can now recover the framework for conformal transformations in this more general framework.
One can in fact define a generator of conformal transformations to be the vector field $\xi= \xi^\mu\del_\mu$ such that  one has
$$
\del_\al\xi^\al g_{\mu\nu}
=\del_\mu \xi^\al g_{\al\nu}+\del_\nu \xi^\al g_{\al\mu} 
\fn$$
which makes a global sense when $C$ is trivial, i.e.~when $M$ is parallelizable as, e.g., when $M=\R^m$.

On the contrary, the splitting \ShowLabel{XiSplit} makes always a global sense.
It is a fortunate situation of having a gauge and a natural structure at the same time which makes the theory peculiar
and allows one to canonically define the action of Weyl transformations and spacetime diffeomorphisms on fields.
Of course this is not the case in a general gauge natural theory and this is when conformal theories become peculiar.

One then defines a field theory by selecting a dynamics with the Lagrangian \ShowLabel{Lag}
that is both generally covariant and Weyl covariant in this precise sense.
In any event, this is a gauge natural theory for the gauge group $\R$ and one can apply the general framework for conservation laws.
The superpotential for the symmetry generator $\Xi$ is then
$$
U=  \Frac[2/3]\Big\{ \( 6 \na^{[\la} R^{\mu]}{}_\ep - \na^{[\la} R \de^{\mu]}_\ep \) \xi^\ep 
+ \( R g^{\nu[\mu} \de^{\la]}_\ep + 6 R^{\nu[\la} \de^{\mu]}_\ep + 3 R_\ep{}^{\nu\la\mu}\) \na_\nu \xi^\ep \Big\} d \si_{\la\mu}
\fn$$
where $d \si_{\la\mu}$ is the local standard basis for $(m-2)$-forms induced by coordinates.

One can see that the superpotential is not dependent on $\ze$ and in fact it depends only on the natural lift of the spacetime vector field $\xi$. 
In this sense it is independent of Weyl transformations.

\NewSection{Hole argument and the physical states}

The {\it hole argument} was originally discussed by Einstein in terms of boundary problems and by Hilbert in terms of initial condition problems; see \ref{Norton}, \ref{Stachel}. We here argue that, when considered in terms of the Cauchy problem, the hole argument poses constraint in the definition of the physical state; see \ref{ADM}.

In order to have a well-posed Cauchy problem one needs to start with a {\it globally hyperbolic} spacetime, i.e.~a manifold $M$ which can be foliated over $\R$. That amounts to require existence of the so--called {\it ADM fibration} $\tau:M\arr \R$.
Since $\R$ is contractible the ADM bundle is trivial and $M\simeq \R\times \Si$ for some 3d manifold $\Si$ describing the {\it space}.

The Cauchy problem is defined by giving initial conditions on a fibre $\tau^{-1}(t_0)=\Si_{t_0}\simeq \Si$, which is called a {\it Cauchy surface}. 
Field equations split in a well-posed Cauchy problem and in an elliptic constraint; for any initial condition satisfying constraints, then the Cauchy problem fixes uniquely the evolution of fields.
Doing that some of the fields are uniquely determined, some are left undetermined. 
The typical situation in GR and gauge theories is having a description of the system which is overdetermined (because of constraints)
and underdetermined (because of the fields left undetermined) at the same time.

The undetermined fields are the mathematical description of our gauge freedom in describing the physical system.
They are the price we have to pay for a gauge covariant description of physics.
For any choice of these undetermined fields one can build a solution of the original (gauge covariant) field equations, which thence is not unique.

The hole argument pin points this non-uniqueness of solutions directly. We shall hereafter state the hole argument for Weyl transformations.
By minor modifications that applies to any other gauge symmetry as well as general covariance.
If one considers a Weyl transformation $\Phi$ which restricts to the identity in a neighbourhood $D$ of a Cauchy surface $\Si_0$,
then called a {\it Cauchy transformation},
and a solution $\si$, then $\si'=\Phi_\ast \si$ is a solution as well (being $\Phi$ a symmetry) and it preserves the initial conditions on $\Si_0$
(since $\Phi$ is the identity around $\Si_0$). Then the solution of field equations is not uniquely determined by initial conditions.

Being this situation quite ubiquitous in physics it seems that determinism is condemned: knowing the world today at the initial condition does not allow to predict its description at different time uniquely, and this is directly linked to gauge symmetries.
Every time we have a symmetry group in which there exist transformations which can be the identity within $D$ and non-trivial out of  $D$ (which is what we call {\it gauge symmetries}) then solutions are not unique.

There is only two possible ways out (beside giving up determinism).
First, theories with gauge symmetries are forbidden or, second, 
the configurations $\si$ and $\si'$ describe the same physical state.

In order to pursue the second possibility and take it seriouosly, one needs in fact to define a group of transformations $\calG$
which preserves the physical state (so that two configurations $\si$ and $\si'$ represent the same physical state iff there exists 
$\Phi\in \calG$ such that $\si'=\Phi_\ast \si$).

Then if $\calG$ is a subgroup of the symmetry group $\calS$, field equations are compatible with the quotient induced by $\calG$
and they induce equations for the physical state which is defined as equivalence classes of configurations with respect to the action 
of $\calG$.
Moreover, if $\calG$ contains all Cauchy transformations the equations induced on the quotient space for the physical state are deterministic
solving the hole argument at the level of physical state if not at the level of configurations.

By quotienting out all Cauchy transformations, we mean at least the group $\calD$ generated by Cauchy transformations since the relation defining the physical system needs to be an equivalence relation. In other words, one needs to define $\calG$ so that
$$
\calD\subset \calG\subset \calS
\fn$$

We shall hereafter argue that for Weyl transformations one has $\calD=\calS$ so that the only option is $\calG=\calS$ and hence 
two configurations differing by a Weyl transformation do define the same physical state.
Our proof does not extend to general diffeomorphisms; see \ref{PhysState}.

Let us prove that any Weyl transformation $\Phi$ on the configuration bundle $C$ or, equivalently, on the structure bundle $P$ is the composition of two Cauchy transformations.
Let us fix an ADM structure on spacetime and consider fibered coordinates $(t, x^i)$. The Weyl transformation $\Phi$,
being a vertical transformation in the structure bundle $P$,
reads as
$$
x'=x
\qquad
l'= \om(t, x) + l
\fn$$
Let us consider two smooth functions $\vp_\pm:\R\arr \R$, called {\it step-functions}, that obey the following properties

\itemitem{i)} $\vp_\pm(t)\ge 0$;

\itemitem{ii)} $\vp_+(t)+ \vp_-(t)=1$;

\itemitem{iii)} $\vp_+(t)= 1$ for $t>1$ and $\vp_+(t)= 0$ for $t<-1$;

By using the step functions we can split the Weyl transformation $\Phi$ as
two Weyl transformations $\Phi_\pm$ defined as
$$
x'=x
\qquad
l'= \vp_\pm(t) \cdot \om(t, x) + l
\fn$$

As a consequence of the properties of step-functions, the transformations $\Phi_\pm$ are Cauchy transformations.
In fact, $\Phi_+$ is the identity in the region $M_-= \tau^{-1}(-\infty, -1)$ (which is in fact a neighbourhood of some Cauchy surface, e.g.~
$\Si_-=\tau^{-1}(-2)$),  $\Phi_-$ is the identity in the region $M_+= \tau^{-1}(1, +\infty)$ and $\Phi=\Phi_+\circ \Phi_-$.

\ms
Since any Weyl transformation sends a physical state into itself then two conformal metrics are the same physical state in conformal gravity.
Accordingly, conformal gravity has less physical states than standard GR. 
In fact, the Ricci scalar transforms under Weyl transformations as
$$
\vp \tilde R= R-\Frac[3/\vp]\Dal \vp + \Frac[3/2\vp^2] \na_\ep \vp \na^\ep \vp
\fn$$

Accordingly, one can easily find a conformal factor $\vp$ for which $R$ and $\tilde R$ are different.
Since the scalar curvature is a diffeomorphism invariant, this shows that $\tilde g$ and $g$ are two configurations which are conformally equivalent though not diffeomorphic equivalent.
Then the two metrics $g$ and $\tilde g$ represents the same physical state in conformal gravity, but two different states in standard GR.

\NewSection{Conclusions and perpectives}

We here considered conformal gravity as a gauge natural theory and discussed conservation laws.
Although we confirm that Noether currents with respect to Weyl transformations identically vanish off-shell as
claimed in a simpler case in \ref{Jackiw}, we also argue that nevertheless conformal gravity is not equivalent to standard GR 
as far as the definition of physical state is concerned.

Further investigations are needed to highlight the meaning of the vanishing of Noether currents and to clarify the physical meaning of the conserved quantities in conformal gravity, (which are in fact conformally invariant quantities).

\Acknowledgements

This paper is dedicated to the memory of Mauro Francaviglia.
We are grateful to Marco Ferraris and Guido Magnano for discussions and comments.

We acknowledge the contribution of INFN (Iniziativa Specifica QGSKY), 
the local research project {\it Metodi Geometrici in Fisica Matematica e Applicazioni} (2015) of Dipartimento di Matematica of University of Torino (Italy). 
This paper is also supported by INdAM-GNFM.

\ShowBiblio

\end